\documentclass[prb,twocolumn,superscriptaddress,nobibnotes]{revtex4}

\usepackage{graphicx}
\usepackage{dcolumn}
\usepackage{bm}
\usepackage{times,mathptmx}
\usepackage{color}
\usepackage{multirow}
\usepackage{enumitem}
\usepackage{chngpage}
\usepackage{amsmath}

\begin{document}


\title{\bf Lithium niobate photonic-crystal electro-optic modulator}

\author{Mingxiao Li}
\affiliation{Department of Electrical and Computer Engineering, University of Rochester, Rochester, NY 14627}
\author{Jingwei Ling}
\affiliation{Department of Electrical and Computer Engineering, University of Rochester, Rochester, NY 14627}
\author{Yang He}
\affiliation{Department of Electrical and Computer Engineering, University of Rochester, Rochester, NY 14627}
\author{Usman A. Javid}
\affiliation{Institute of Optics, University of Rochester, Rochester, NY 14627}
\author{Shixin Xue}
\affiliation{Institute of Optics, University of Rochester, Rochester, NY 14627}
\author{Qiang Lin}
\email{qiang.lin@rochester.edu}
\affiliation{Department of Electrical and Computer Engineering, University of Rochester, Rochester, NY 14627}
\affiliation{Institute of Optics, University of Rochester, Rochester, NY 14627}



\begin{abstract}
Modern advanced photonic integrated circuits require dense integration of high-speed electro-optic functional elements on a compact chip that consumes only moderate power. Energy efficiency, operation speed, and device dimension are thus crucial metrics underlying almost all current developments of photonic signal processing units. Recently, thin-film lithium niobate (LN) emerges as a promising platform for photonic integrated circuits. Here we make an important step towards miniaturizing functional components on this platform, reporting probably the smallest high-speed LN electro-optic modulators, based upon photonic crystal nanobeam resonators. The devices exhibit a significant tuning efficiency up to 1.98~GHz/V, a broad modulation bandwidth of 17.5~GHz, while with a tiny electro-optic modal volume of only 0.58~$\mu {\rm m}^3$. The modulators enable efficient electro-optic driving of high-Q photonic cavity modes in both adiabatic and non-adiabatic regimes, and allow us to achieve electro-optic switching at 11~Gb/s with a bit-switching energy as low as 22 fJ. The demonstration of energy efficient and high-speed electro-optic modulation at the wavelength scale paves a crucial foundation for realizing large-scale LN photonic integrated circuits that are of immense importance for broad applications in data communication, microwave photonics, and quantum photonics.

\end{abstract}

\maketitle

\section*{Introduction}

High-speed electro-optic modulation underlies many important applications ranging from optical communication \cite{Wooten00}, microwave photonics \cite{Capmany19}, computing \cite{Vladimir15}, frequency metrology \cite{Diddams18}, to quantum photonics \cite{Smith17}. A variety of approaches have been employed for electro-optic modulation, such as carrier plasma dispersion \cite{Reed10, Keeler18}, electro-absorption \cite{Nakano18, Michel08}, and Pockels effect \cite{Wooten00, Koos15}, the latter of which is particularly interesting since the Pockels effect offers an ultrafast and pure refractive-index modulation over an extremely broad optical spectrum while without introducing extra loss. The best-known electro-optic Pockels material is probably lithium niobate (LN), which has been widely used in telecommunication \cite{Wooten00}. Recently, thin-film monolithic LN \cite{Gunter12, Bowers18} emerges as a promising platform, where low-loss and high-quality photonic integration together with the strong Pockels effect enables superior modulation performance \cite{Gunter07, Reano13, Reano14, Fathpour15, Fathpour16, Amir17, Loncar18, Prather18, Loncar18_2, Shayan18, Fathpour18, Cai19, Cai19_2, Loncar19}, showing great potential as an excellent medium for photonic integrated circuits and future photonic interconnect. 

\begin{figure*}[t!]
	\centering\includegraphics[width=2.0\columnwidth]{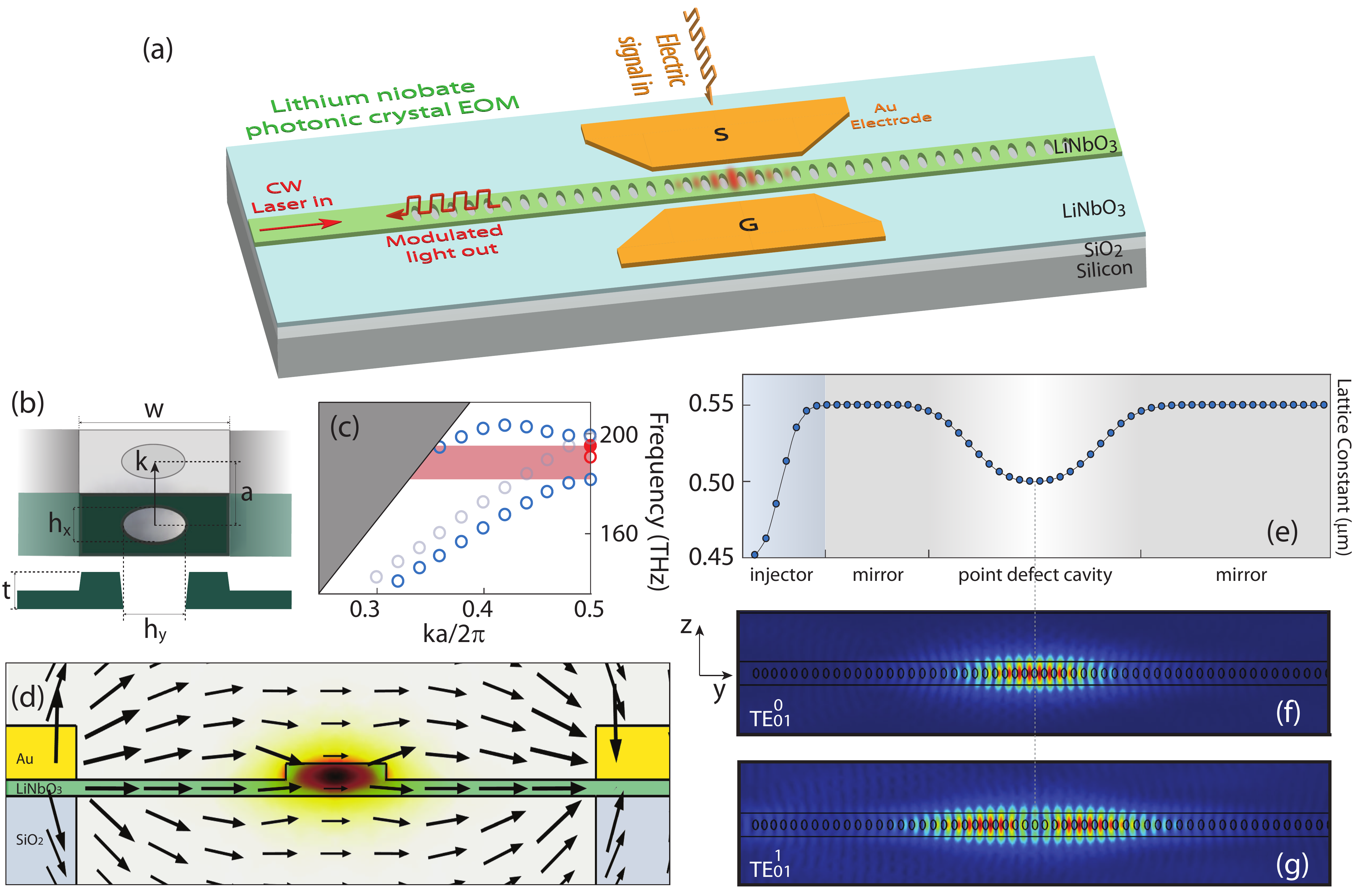}
	\caption{\label{Fig1} Design of LN photonic crystal EOM. {\bf (a)} Schematic of the LN photonic crystal EOM. {\bf (b)} The structure of the unit cell (top: top view; bottom: cross-section view). The LN photonic crystal nanobeam has a width of $w=1200$~nm, layer thickness of $t=300$~nm, and a partially etched wing layer with a thickness of 150~nm. The elliptical hole has dimensions of $h_{x}=270$~nm and $h_{y}=490$~nm, and a fully etched depth of 300~nm. The full cross section is shown in (d). {\bf (c)}  Dispersion property of the partially etched LN photonic crystal nanobeam, simulated by the finite element method (FEM). The blue open circles show the dielectric and air bands. The red solid and open circles denote the fundamental and second-order TE-like cavity modes shown in (f) and (g). Our simulations show that there exhibits another mode with eigenfrequency within the band gap (gray open circles). This mode, however, has only negligible perturbation to the dielectric mode due to distinctive spatial symmetry, thus not affecting the quality of the defect cavity mode. {\bf (d)} Cross-section schematic of the EOM structure, where the arrow profile shows the RF electric field distribution and the color profile shows the optical cavity mode field distribution, both simulated by the FEM method. {\bf (e)} Lattice constant as a function of position, which is optimized for low insertion loss together with high radiation-limited optical Q. {\bf (f)} Top view of the FEM-simulated optical mode field profile of the fundamental TE-like cavity mode $TE^0_{01}$. The left inset shows the orientation of the LN crystal where the optical axis is along the $z$ direction. {\bf (e)} Simulated optical mode field profile of the second-order TE-like cavity mode $TE^1_{01}$.}
\end{figure*}

Power efficiency is crucial for the application of electro-optic modulator (EOM), which depends sensitively on the physical size of the device \cite{Miller17}. Scaling an EOM down to a small footprint would reduce the device capacitance and thus decrease the switching energy \cite{Miller17, Sorger15}, which is indispensable for all practical applications. Among various device geometries, photonic crystal nanoresonators are particularly beneficial in this regard, given their exceptional capability of controlling light confinement and light-matter interactions on the sub-wavelength scale. In the past decade, photonic-crystal EOMs have been developed on various material platforms such as silicon \cite{Notomi09, Baba12, Notomi14}, GaAs  \cite{Jelena11}, InP \cite{Notomi19}, polymers \cite{Krauss09, Chen16}, ITO \cite{AlanX19}, etc. For lithium niobate, however, the EOMs developed so far \cite{Wooten00, Gunter07, Reano13, Reano14, Fathpour15, Fathpour16, Amir17, Loncar18, Prather18, Loncar18_2, Shayan18, Fathpour18, Cai19, Cai19_2, Loncar19} generally exhibit significant dimensions, leading to significant power required to drive the EOMs. Although attempts have been made to explore the electro-optic effect in LN photonic crystals \cite{Bernal12, Bernal12_3, Bernal14}, the low device quality and poor optoelectronic integration unfortunately limit seriously the operation speed. To date, it remains an open challenge in realizing a high-speed and energy-efficient modulator at the wavelength scale on the monolithic LN platform. 

\begin{figure*}[htbp]
	\centering\includegraphics[width=1.40\columnwidth]{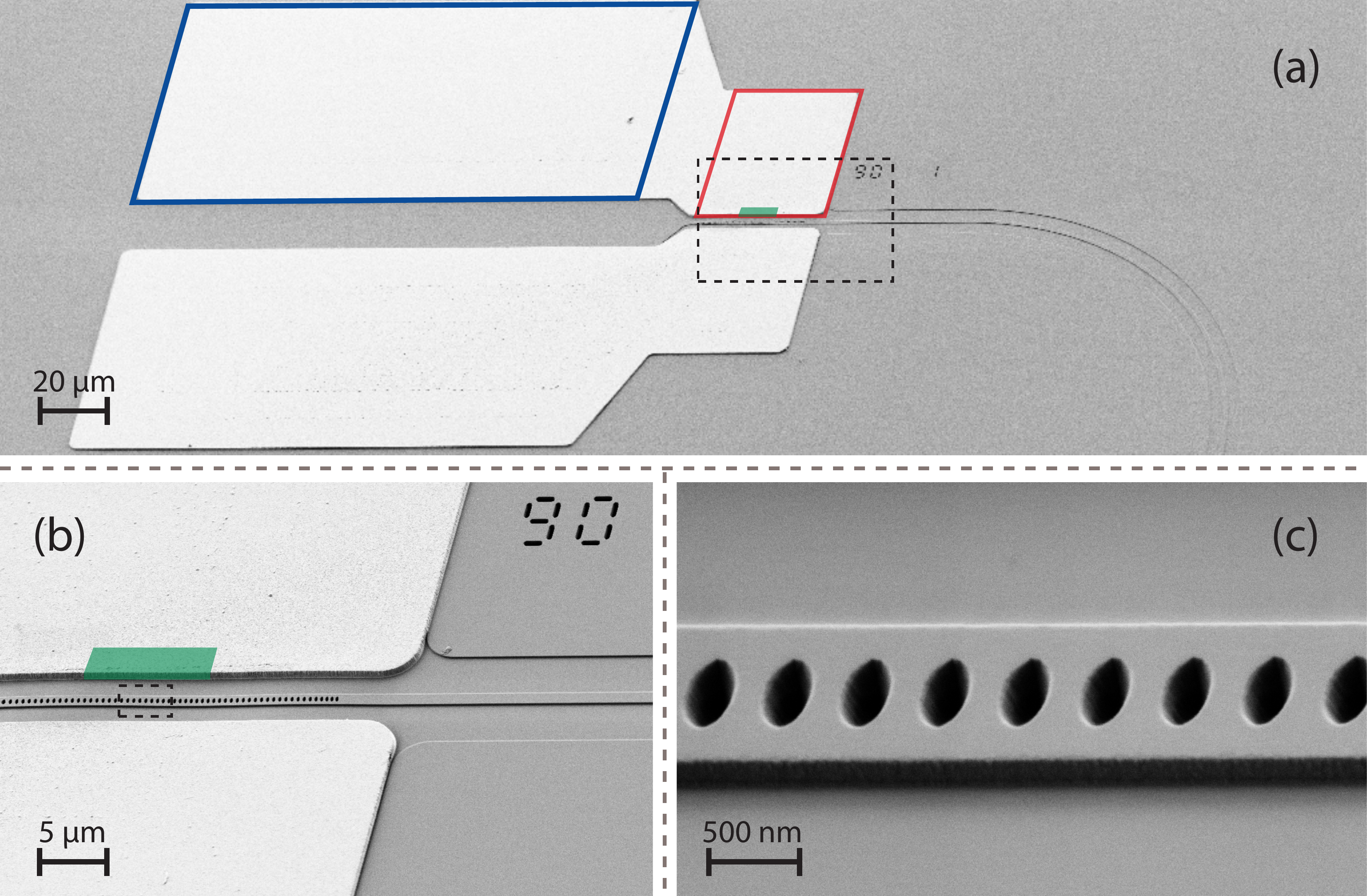}
	\caption{\label{Fig2} Scanning electron microscopic (SEM) image of a fabricated EOM device. {\bf (a)} Full SEM image of the whole device structure. The region highlighted in red is the electrode used to drive the photonic crystal nanoresonator. That highlighted in blue indicates the large metal pad used for contacting the RF probe. The green region indicates the electrode that can be shrunk to in the future design. {\bf (b)} Zoom-in image of the photonic crystal resonator and electrodes, corresponding to the dashed rectangular region in (a). {\bf (c)} Further zoom-in image showing the detailed structure of the photonic crystal defects cavity, corresponding to the dashed rectangular region in (b). }
\end{figure*}

Here we report high-speed and energy-efficient LN photonic crystal EOMs, which exhibits a tiny electro-optic modal volume of only $\sim$ 0.58~$\mu{\rm m}^3$, the smallest among all high-speed LN EOMs ever reported \cite{Wooten00, Gunter07, Reano13, Reano14, Fathpour15, Fathpour16, Amir17, Loncar18, Prather18, Loncar18_2, Shayan18, Fathpour18, Cai19, Cai19_2, Loncar19}. The sub-wavelength-scale EOM cavity enables compact optoelectronic integration to achieve not only a high electro-optic tuning efficiency up to 16~pm/V (corresponding to 1.98~GHz/V) that is significantly beyond other LN EOM resonators \cite{Gunter07, Reano13, Reano14, Fathpour15, Amir17, Loncar18, Fathpour18, Loncar19}, but also a large modulation bandwidth up to 17.5~GHz that reaches the photon-lifetime limit of the EOM cavity. The fully on-chip design achieves a full-swing extinction ratio of 11.5~dB. With these devices, we are able to realize efficient driving of the optical mode in both adiabatic sideband-unresolved and non-adiabatic sideband-resolved regimes, and to observe the transition in between. As an example application, we demonstrate electro-optic switching of non-return-to-zero signal at a rate of 11~Gb/s, with a switching energy as low as 22~fJ/bit that is more than one order of magnitude smaller than other LN EOMs  \cite{Wooten00, Gunter07, Reano13, Reano14, Fathpour15, Fathpour16, Amir17, Loncar18, Prather18, Loncar18_2, Shayan18, Fathpour18, Cai19, Cai19_2, Loncar19}. To the best of our knowledge, this is the smallest LN EOM ever demonstrated with combined high speed and energy efficient operation. 

\section*{Device design and fabrication}

Recently, there have been significant advance in high-Q LN photonic-crystal nanoresonators \cite{Liang17, Li19, Amir19, Mingxiao192}, which led to the demonstration of intriguing phenomena and functionalities such as photorefraction quenching \cite{Liang17}, harmonic generation \cite{Li19}, piezo-optomechanics \cite{Amir19}, and all-optical resonance tuning \cite{Mingxiao192}. 

For EOM, We adopt one-dimensional photonic-crystal nanobeam as the basic underlying structure (Fig.~\ref{Fig1}(a)) since it supports compact optical and electrical integration to enhance the electro-optic response. Due to the high permittivity of LN at radio frequency, the commonly used full surrounding air cladding \cite{Liang17, Mingxiao192, Amir19} is not suitable for EOM since it would significantly reduce the coupling between the optical and electric fields. To maximize the electro-optic interaction, we utilize a partially etched structure with a rib-waveguide-like cross section, leaving a 150-nm-thick wing layer for the electrodes to sit on (Fig.~\ref{Fig1}(a) and (d)). Although the breaking of the mirror symmetry along the normal direction of the device plane considerably alters the band gap of the photonic crystal (Fig.~\ref{Fig1}(c)), optimization of the photonic potential via an appropriate pattern of lattice constant (Fig.~\ref{Fig1}(e)) is still able to produce a well-confined point-defect cavity, with a simulated optical Q of $\sim 10^5$ for the fundamental transverse-electric-like (TE-like) cavity mode, $TE^0_{01}$, shown in Fig.~\ref{Fig1}(f). The cavity mode exhibits an extremely small electro-optic modal volume of $1.52 (\lambda/n)^3 \sim0.58~\mu {\rm m}^3$ (where $n$ is the refractive index of LN), which is the smallest among all LN EOMs ever reported \cite{Wooten00, Gunter07, Reano13, Reano14, Fathpour15, Fathpour16, Amir17, Loncar18, Prather18, Loncar18_2, Shayan18, Fathpour18, Cai19, Cai19_2, Loncar19}, to the best of our knowledge. 

The photonic crystal cavity is oriented such that the dominant optical field is in parallel with the optical axis of underlying LN medium (Fig.~\ref{Fig1}(f)), so as to take advantage of the largest electro-optic component $r_{33}$ of lithium niobate. The electrodes are designed to be placed close to the photonic-crystal resonator (Fig.~\ref{Fig1}(d)) to maximize the in-plane electric field ${E_z}$, while preventing potential loss induced by metal absorption. The compact device structure design results in a significant electro-optic tuning efficiency of 1.81~GHz/V, simulated by the finite element method. The electrodes are designed to have a length of 30~$\mu$m to ensure a full coverage of the applied electric field over the entire photonic crystal structure. Numerical simulations show that the device exhibits a small capacitance C of $C$ = $\sim$22~fF, which is more than one order of magnitude smaller than other LN EOMs \cite{Wooten00, Gunter07, Reano13, Reano14, Fathpour15, Fathpour16, Amir17, Loncar18, Prather18, Loncar18_2, Shayan18, Fathpour18, Cai19, Cai19_2, Loncar19}. Therefore, we expect our devices to have much higher energy efficiency, as will be shown in the following sections. 

\begin{figure*}[htbp]
	\centering\includegraphics[width=1.80\columnwidth]{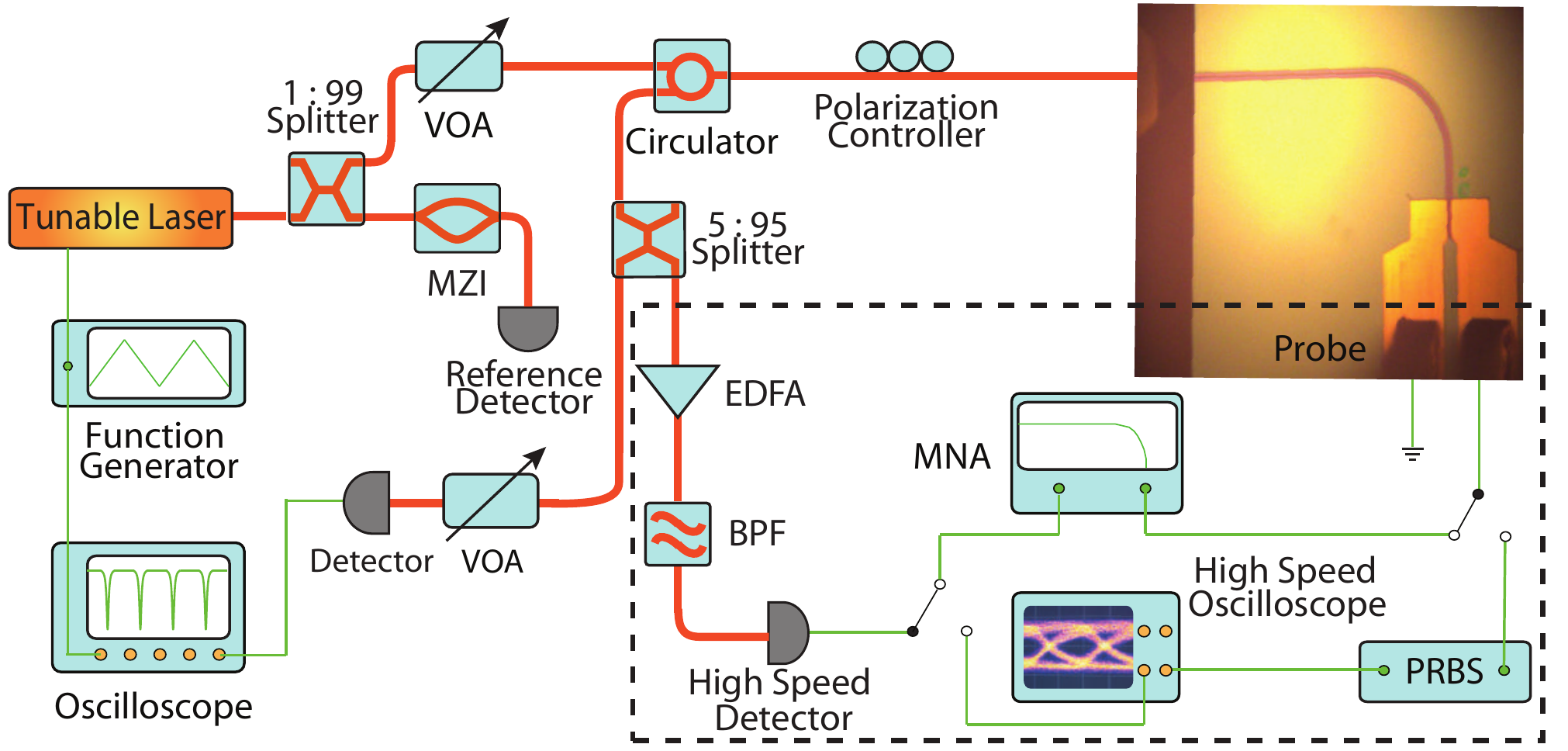}
	\caption{\label{Fig3} Experimental testing setup. Light is coupled into and out of the EOM chip via one lensed fiber. The inset shows an optical microscopic image of an EOM with the RF probe in contact. The equipment in the highlighted dashed box is used for characterizing the performance of electro-optic modulation. VOA: variable optical attenuator; MZI: Mach-Zehnder interferometer; EDFA: erbium doped fiber amplifier; BPF: band pass filter; MNA: microwave network analyzer;  PRBS: pseudo-random binary sequence source}
\end{figure*}

For simplicity of testing, the EOM is designed such that light is coupled into and out of the EOM via only one side of the cavity (Fig.~\ref{Fig1}(a)). As such, the photonic-crystal mirror on the right side of the defect cavity is designed to be of 100\% reflection, while that on the left side has decreased number of holes (Fig.~\ref{Fig1}(e)) to enable a partial reflection/transmission, with the hole number optimized for a critical coupling to the cavity. To support on-chip integration, light is coupled to the EOM cavity via an on-chip waveguide (Fig.~\ref{Fig1}(a)), where an injector section (Fig.~\ref{Fig1}(e)), with the lattice constant varying from 450 to 550~nm, is designed and placed in front of the left mirror to reduce the coupling loss. 

The devices were fabricated on a $300$-nm-thick x-cut single-crystalline LN thin film bonded on a 3-${\mu}$m silicon dioxide layer sitting on a silicon substrate (from NanoLN). The photonic crystal hole structure was patterned with ZEP-$520$A positive resist via electron-beam lithography, which was then transferred to the LN layer with an Ar$^+$ plasma milling process to etch down the full 300-nm depth. The resist residue was removed by a further O$^+$ plasma etching. A second exposure is then performed to define the waveguide structure, which is partially etched by 150~nm with the same process. After the residue removal, we used diluted hydrofluoric acid to undercut the buried oxide layer to form a suspended photonic crystal membrane structure [Fig. \ref{Fig1}(d)]. The metal electrode layer (10 nm Ti/500 nm Au) was deposited by an electron-beam evaporator and the electrode structure was formed by a lift-off process via ZEP-$520$A.

Figure \ref{Fig2} shows a fabricated device. The large metal pads (highlighted in blue box) are used simply as the contacts for the air-coplanar probe (Formfactor Acp65-A-GSG-100) for applying the RF driving signal (see also the inset of Fig.~\ref{Fig3}). The impedance of the metallic structure is optimized to minimize the coupling loss of the RF signal from the pads to the device. The high quality of device fabrication as indicated by the device images implies high performance of the EOM, as we will show below.

\section*{Device characterization and electro-optic properties}

\begin{figure}[htbp]
	\centering\includegraphics[width=1.0\columnwidth]{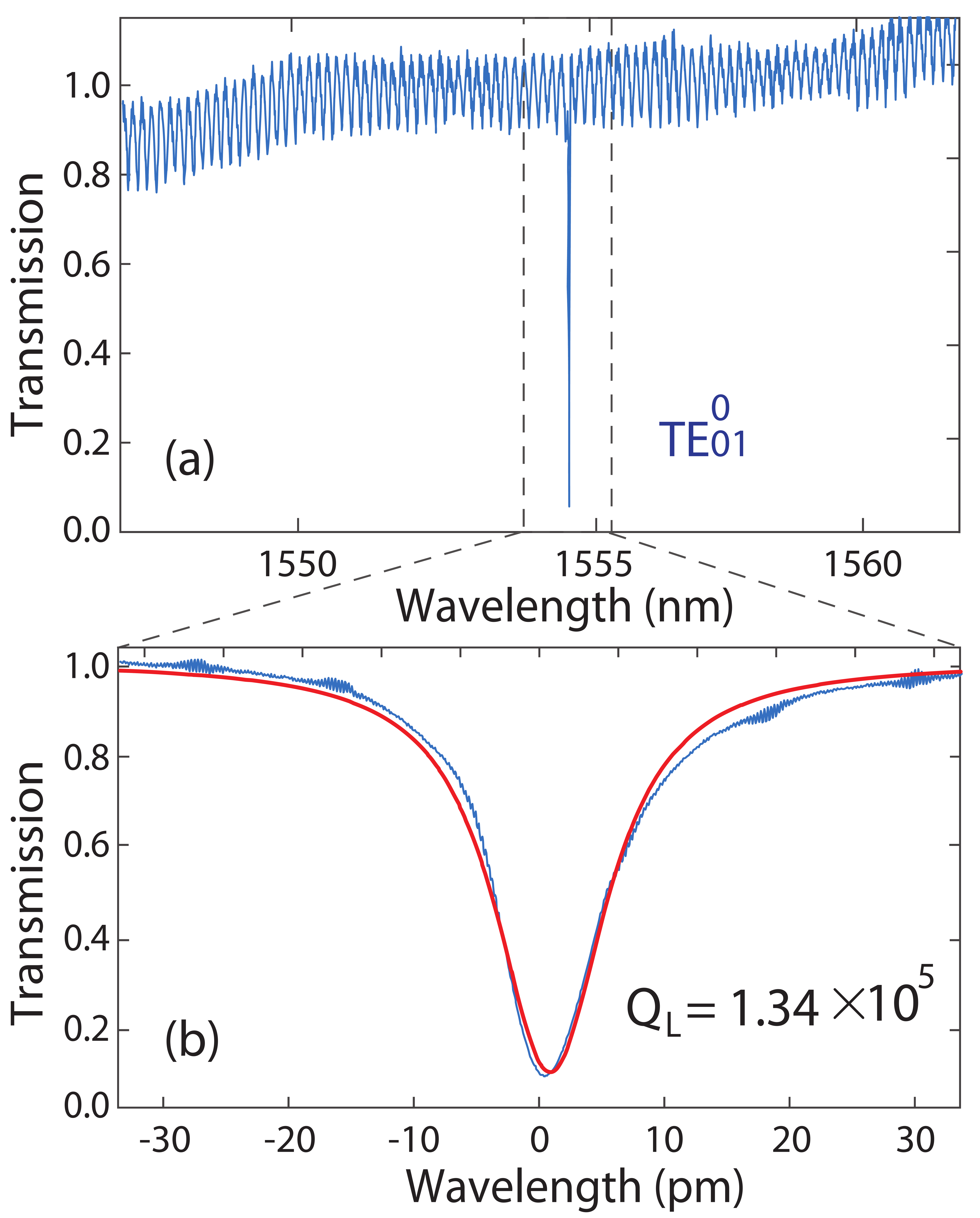}
	\caption{\label{Fig4} Linear optical property of a fabricated LN photonic crystal EOM. ({\bf a}) Laser-scanned transmission spectrum in the telecom band. ({\bf b}) Detailed transmission spectrum of the fundamental TE-like cavity mode $TE^0_{01}$ at wavelength of 1554.47~nm, with the experimental data shown in blue and the theoretical fitting shown in red. }
\end{figure}

\begin{figure}[b!]
	\centering\includegraphics[width=1.0\columnwidth]{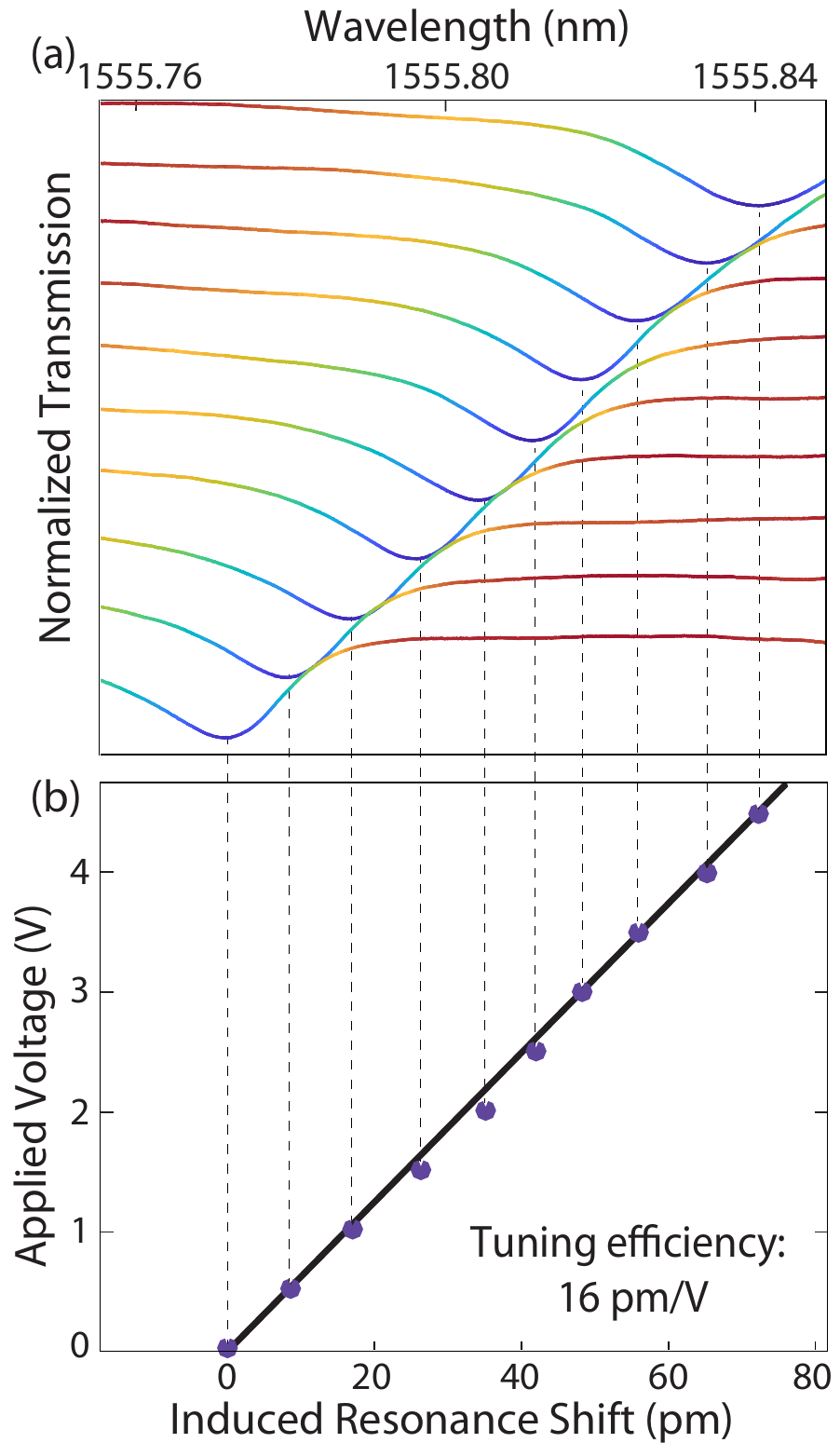}
	\caption{\label{Fig5}  Electro-optic tuning property of an LN photonic crystal EOM. {\bf (a)} Recorded transmission spectrum of the EOM cavity as a function of applied DC voltage from 0 to 4.5~V, with a voltage step of 0.5~V. {\bf (b)} Recorded resonance shift as a function of applied DC voltage, where the experimental data are shown in black dots and the blue line is a linear fitting to the data. }
\end{figure}

To characterize the optical and electro-optic properties of the devices, a continuous-wave tunable laser (Santec TSL-510) was launched onto the chip via a lensed fiber. The light reflected from the EOM was collected by the same lensed fiber, routed by a circulator, and then delivered to a photodiode for detection. Figure \ref{Fig3} illustrates the schematic of the experimental testing setup, where the inset shows an optical image of the device with the RF probe in contact. The insertion loss from the on-chip coupling waveguide to the photonic-crystal cavity is measured to be around 2.2~dB, calibrated by subtracting the coupling loss from the facet and circulator transmission loss. To characterize the performance of high-speed modulation, the majority of the modulated light output was amplified by an erbium-doped fiber amplifier to boost the power, passed through a bandpass filter to remove the amplifier noise, and was then detected by a high-speed detector (New Focus 1024). The detector output was recorded either by a microwave network analyzer (Keysight N5235B) for characterizing the modulation bandwidth or by a sampling oscilloscope module (Keysight 54754A) to record the eye diagram of the switching signal.

Figure \ref{Fig4}(a) shows the transmission spectrum of an EOM when the laser is scanned in the telecom band. The device exhibits a resonance at 1554.47~nm, which corresponds to the fundamental TE-like cavity mode $TE^0_{01}$ (Fig.~\ref{Fig1}(f)). As shown in Fig.~\ref{Fig4}(b), the $TE^0_{01}$ mode exhibits a high loaded optical Q of $1.34\times10^5$, which is very close to our numerical simulation, indicating the negligible impact of the electrodes on the optical quality. The cavity resonance exhibits a coupling depth of 93\%, corresponding to a full-swing extinction ratio of 11.5~dB. This value can be improved in the future by further optimizing the partially reflective photonic-crystal mirror (Fig.~\ref{Fig1}(e)). The device also exhibits a second-order TE-like cavity mode $TE^1_{01}$ (Fig.~\ref{Fig1}(g)) at 1604.13 nm (not shown) with a loaded optical Q of $3.03\times10^4$. 

To show the electro-optic tuning property, we applied a DC voltage to the chip and monitored the cavity transmission spectrum of the $TE^0_{01}$ mode. As shown in Fig.~\ref{Fig5}(a), the cavity resonance tunes smoothly with the applied voltage, without any degradation to the lineshape or coupling depth, clearly showing the pure dispersive electro-optic tuning as expected from the Pockels effect. We have applied a voltage of 25~Volts to the device (not shown in the figure) and did not observe any degradation. Figure \ref{Fig5}(b) shows a clear linear dependence of the induced resonance wavelength shift on the applied voltage, from which we obtained a tuning slope of 16~pm/V (corresponding to a frequency tuning slope of 1.98~GHz/V), close to our design. This value is significantly larger than those in other LN EOM resonators \cite{Gunter07, Reano13, Reano14, Fathpour15, Amir17, Loncar18, Fathpour18, Loncar19}, which is primarily benefited from the strong optical field confinement, large optical and electric field overlap, and the resulting compact optical and electric integration offered by our devices. 


\section*{Electro-optic modulation}

\begin{figure*}[t!]
	\centering\includegraphics[width=2.0\columnwidth]{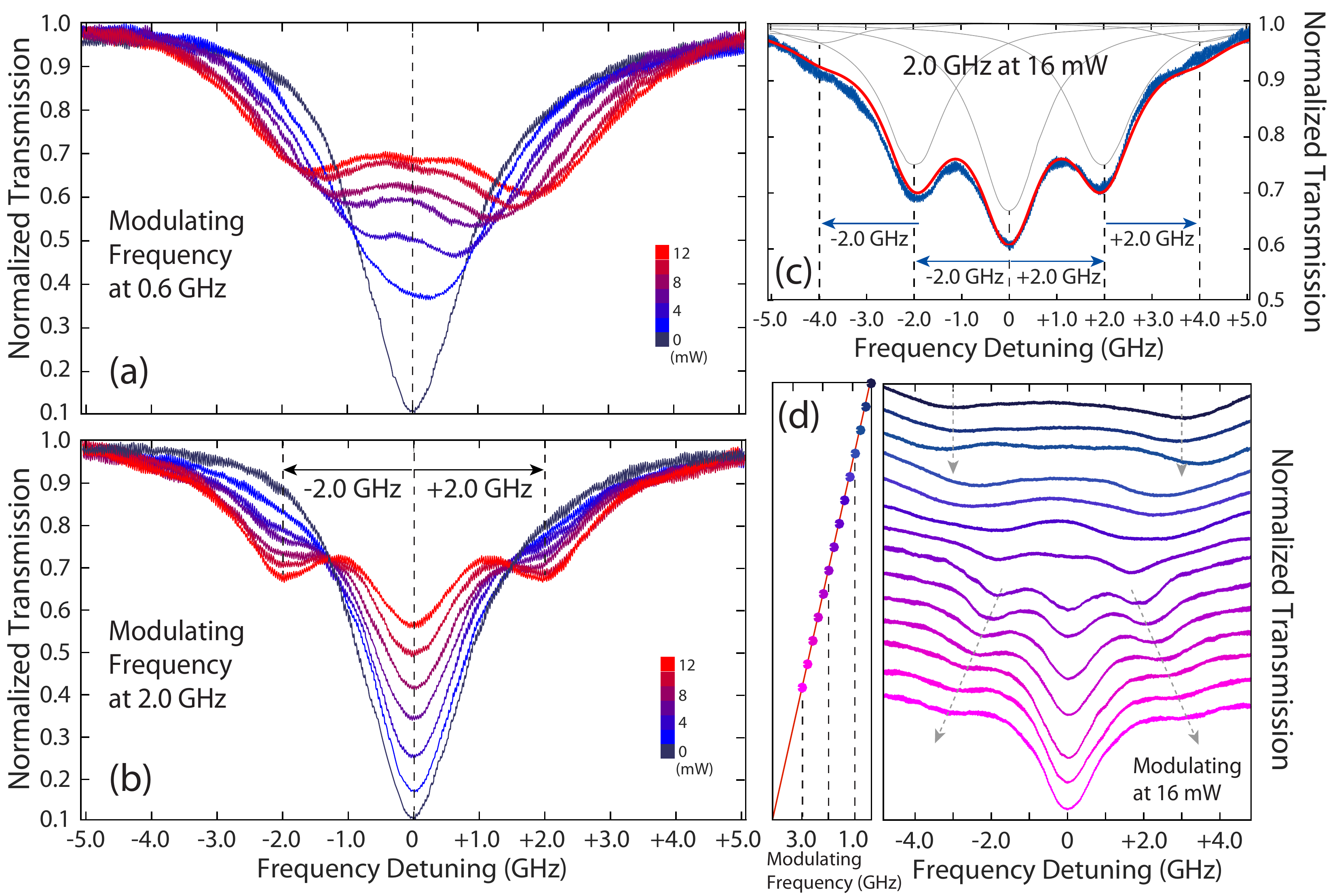}
	\caption{\label{Fig6} Electro-optic modulation of a high-Q optical cavity resonance. {\bf (a)} Recorded transmission spectra of the $TE^0_{01}$ cavity mode with RF driving signal at 7 different powers from 0 to 12~mW, with a power step of 2~mW, modulated at 0.6~GHz. {\bf (b)} Same as (a) but with a modulation frequency of 2.0~GHz. {\bf (c)} Detailed spectrum with RF driving signal at 2.0 GHz with power of 16 mW. The gray lines are in Lorentzian shapes and the sum of all gray lines are showed in red lines fitted by the theory. The dashed lines note the spacing of the sidebands. {\bf (d)} Recorded transmission spectra at different RF modulation frequency from 0.4 to 3.0~GHz, with a frequency step of 0.2~GHz. The RF driving power is 16 mW. }
	
\end{figure*}

The high efficiency of electro-optic tuning together with the high optical quality of the EOM resonator enables efficient electrical driving of the optical mode into different dynamic regimes. To show this phenomenon, we applied a sinusoidal RF signal at a certain frequency to the EOM and monitored the transmission spectrum of the device by scanning laser back and forth across the cavity resonance. The laser wavelength is scanned at a repetition rate of $\sim$15~Hz, so we primarily monitored the time-averaged cavity transmission. 

When the EOM is driven at a modulation frequency of 600~MHz much smaller than the cavity linewidth of 1.4~GHz, increasing the driving power simply broadens the transmission spectrum into one with two shallow side lobes, as shown in Fig.~\ref{Fig6}(a), with a broadened spectral linewidth dependent on the driving power. This is a typical signature of resonance modulation in the sideband-unresolved regime, where the cavity resonance follows adiabatically the electric driving signal in a sinusoidal fashion, resulting in a broadened average transmission spectrum as shown in Fig.~\ref{Fig6}(a). 

When the modulation frequency is increased to 2.0~GHz greater than the cavity linewidth, the cavity is too slow to follow the electro-optic modulation, which results in the frequency conversion of photons into sidebands with frequency separation equal to the modulation frequency. Consequently, the transmission spectrum transforms into a multi-resonance spectrum, as shown in Fig.~\ref{Fig6}(b). Increasing the electrical driving power now does not perturb the positions of the resonance dips, but rather changes their relative magnitudes since the magnitudes of the created sidebands depends on the driving amplitude \cite{Usman19}. This phenomenon is shown more clearly in Fig.~\ref{Fig6}(c), where a driving power of 16~mW (corresponding peak-to-peak driving voltage, ${\rm V_{pp}}$, of ${\rm V_{pp}} =2.5$~V) splits the cavity resonance into five with notable magnitudes (black curve), resulting in a cavity transmission with five side lobes (blue curve).   

Electro-optic modulation enables arbitrary modulation of cavity resonance within the bandwidth allowed by the driving circuit. This is in strong contrast to piezoelectric acoustic modulation which is confined to the vicinity of mechanical resonance frequency \cite{Amir19, Loncar196, Piazza19}. Such flexibility allows us to observe direct transition between the adiabatic driving regime and the non-adiabatic regime simply by continuously sweeping the modulation frequency to across the cavity linewidth. Figure \ref{Fig6}(d) shows an example. When the modulation frequency is below 1.0~GHz, The transmission spectrum remains fairly similar regardless of modulation frequency, as expected from the adiabatic driving discussed above. However, when the modulation frequency is tuned above 1.0~GHz towards the cavity linewidth, the two side lobes moves towards each other and the spectral shape is considerably distorted, until around 1.8~GHz where the transmission spectrum splits into three lobes, with the two side lobes located about 1.8~GHz from the center. Further increase of the modulation frequency shifts apart the two side lobes accordingly, with amplitude decreased, while the position of the center lobe remains unchanged, as expected from the non-adiabatic driving. The flexible electro-optic modulation shown here may offer a convenient method for controlling the spectrotemporal properties of photons inside the cavity and for creating exotic quantum states \cite{Usman19} that are crucial for quantum photonic applications.

\begin{figure*}[htbp]
	\includegraphics[width=2.0\columnwidth]{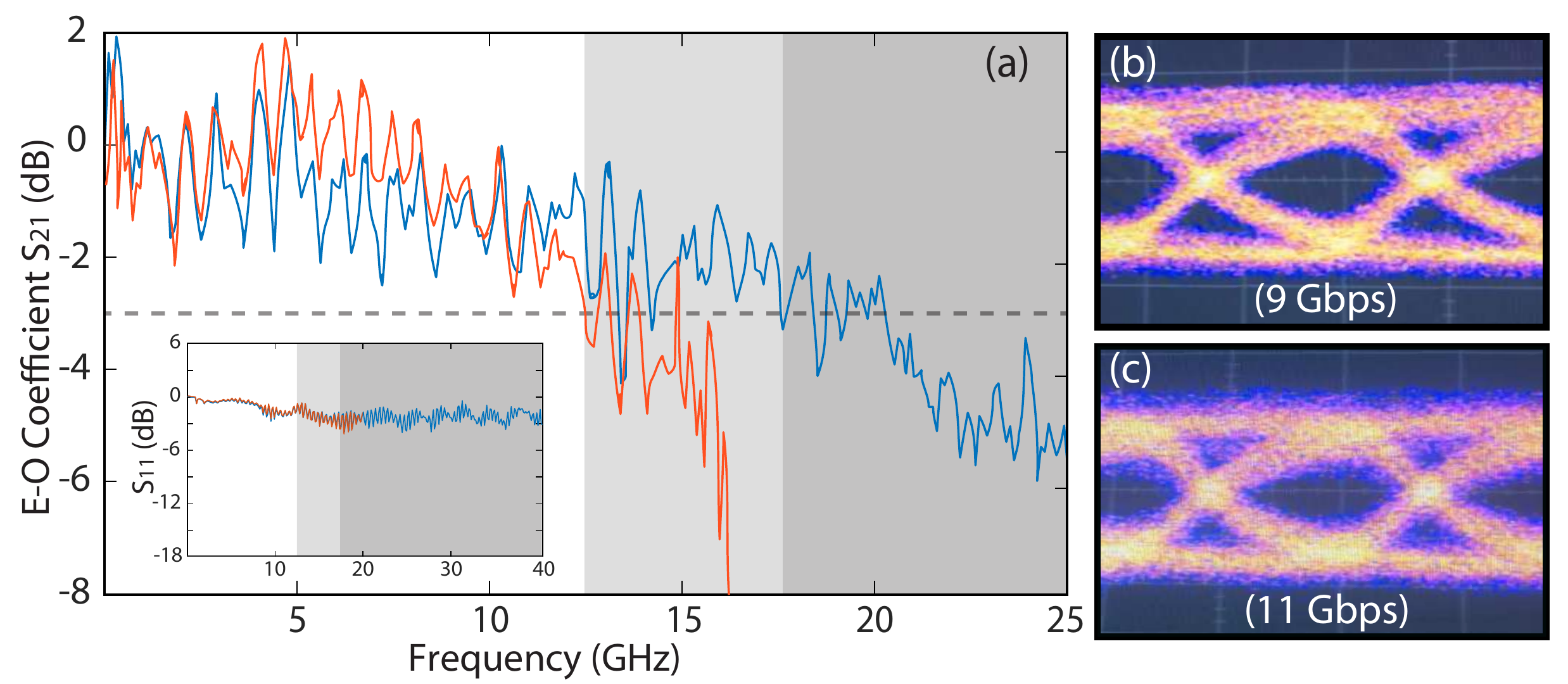}
	\caption{\label{Fig7}  High-speed electro-optic switching. {\bf (a)} Measured scattering parameter $S_{21}$ for one device with Q around 14000 in blue and another with Q around 20000 in orange. The gray dashed line notes the 3 dB cutoff frequency with the gray region represents the bandwidth limit for two device respectively. The inset shows the $S_{11}$ reflection scattering parameter for both devices. {\bf (b) and (c)} Eye diagrams of the photonic crystal EOM, measured with $2^7$-1 NRZ PRBS with a driving voltage of Vpp=2V. }
\end{figure*}

\section*{Electro-optic switching}

The electro-optic modulation demonstrated in the previous section indicates the potential high-speed operation of the EOMs. To show this feature, we selected another similar device on the same chip, which has a lower loaded optical Q of 14000. Figure \ref{Fig7}(a) shows the electro-optic modulation response of the device (blue curve), which exhibits a 3-dB modulation bandwidth up to around 17.5~GHz. This value primarily reaches the photon-lifetime limit of the EOM cavity ($\sim$11~ps), as the electrode circuit has much broader spectral response as indicated by the flat S$_{11}$ reflection spectrum shown in the inset of Fig.~\ref{Fig7}(a). As the modulation bandwidth is primarily related to the optical Q of the device, it can be engineered flexibly for different application purposes, simply by choosing device with appropriate optical Q. The red curve in Fig.~\ref{Fig7}(a) shows another example of a device with optical Q of 20000, which exhibits a 3-dB bandwidth of about 12.5~GHz. 

The broad modulation bandwidth of these devices would thus enable high-speed electro-optic switching. As an example application, we applied non-return-to-zero (NRZ) signal with a ($2^7$-1)-bit pseudo-random binary sequence (PRBS) to an EOM with a ${\rm V_{pp}}$ of 2.0~Volts. Figure \ref{Fig7}(b) and (c) show the recorded eye diagrams at two different bit rates of 9 and 11~Gb/s, respectively, which show clear open eyes. The demonstrated bit rate is currently limited by our PRBS generator (Agilent 70843B) which has a maximum bit rate of 12~Gb/s. However, negligible degradation observed between Fig.~\ref{Fig7}(b) and (c) implies that the EOM could operate at higher bit rates, which will left for future demonstration. The bit switching energy for NRZ signal is given by $\frac{1}{4}CV_{\rm pp}^2$ \cite{Miller17}, which is about 22~fJ/bit in our EOM. This value is the smallest switching energy ever reported for LN EOMs  \cite{Wooten00, Gunter07, Reano13, Reano14, Fathpour15, Fathpour16, Amir17, Loncar18, Prather18, Loncar18_2, Shayan18, Fathpour18, Cai19, Cai19_2, Loncar19}, clearly showing the high energy efficiency of our devices.

\section*{Discussion and conclusion}

The energy efficiency of the LN photonic crystal EOM can be further improved since our current devices are not optimized. For example, the capacitance of our device can be significantly decreased since the majority of the metallic parts in the current devices are used for coupling the RF driving signal, which can be removed in a future on-chip integration design. The 50-$\mu$m width of the electrode (Fig.~\ref{Fig2}, red box) is used primarily for impedance matching to the large metal pad for probe contact, which can be decreased to 3~$\mu$m for a fully on-chip operation \cite{Notomi19}. On the other hand, the 30-$\mu$m length of the electrode is overly conservative since it covers the full length of photonic crystal structure including the injector, mirrors, and the cavity (Fig.~\ref{Fig1}(e) and Fig.~\ref{Fig2}). Essentially, only the 10-$\mu$m long point-defect cavity requires electric driving to achieve electro-optic modulation. Therefore, the electrodes can be shrunk to $10 \times 3~{\mu {\rm m}}^2$, which would reduce the capacitance considerably to $\sim$0.27~fF ($\sim$1.0~fF if including the integrated wires \cite{Notomi19}), according to our FEM simulations. On the other hand, the electrodes are currently placed far from the photonic crystal cavity so as to leave the optical mode intact to achieve a high optical Q. For the application of high-speed electro-optic switching, our simulations show that the electrode-waveguide spacing can be decreased to 1.5~$\mu$m for an optical Q of $\sim$5000 (corresponding to a modulation bandwidth of $\sim$ 45~GHz), which will improve the modulation efficiency to 2.38~GHz/V. We expect that these optimization would significantly improve the energy efficiency of the LN photonic crystal EOM, further decreasing the switching energy down to sub-femtoJoule level. 

In the current EOMs shown above, light is coupled into and out of the EOMs via a same side of the cavity, which is not convenient in practice since a circulator is required to separate the modulated light for the laser input. This can be changed simply by engineering the photonic crystal mirror on the other side to function as the output port. 

In summary, we have demonstrated high-speed LN EOMs with a broad modulation bandwidth of 17.5~GHz, a significant tuning efficiency up to 1.98~GHz/V, and an electro-optic modal volume as small as 0.58~$\mu {\rm m}^3$. We believe this is the first LN EOM ever reported with such combined device characteristics and modulation performance. With these devices, we are able to demonstrate efficient electrical driving of high-Q cavity mode in both adiabatic and non-adiabatic regimes and to observe transition in between. We are also able to achieve high-speed electro-optic switching of at 11 Gb/s, with switching energy as low as 22~fJ/bit. The demonstration of energy efficient and high-speed EOM at the wavelength scale paves an important step for device miniaturization and high density photonic integration on the monolithic LN platform, which is expected to find broad applications in communication, computing, microwave signal processing, and quantum photonic information processing.


\section*{Funding Information} National Science Foundation (NSF) (EFMA-1641099, ECCS-1810169, and ECCS-1842691); the Defense Threat Reduction Agency-Joint Science and Technology Office for Chemical and Biological Defense (grant No.~HDTRA11810047).

\section*{Acknowledgment} The authors thank Professor Hui Wu and Professor Wayne Knox for the use of their equipment. They also thank Wuxiucheng Wang, Lejie Lu, and Ming Gong for valuable discussions and help on testing. This work was performed in part at the Cornell NanoScale Facility, a member of the National Nanotechnology Coordinated Infrastructure (National Science Foundation, ECCS-1542081).



\begin{thebibliography}{99}	
			
			\bibitem{Wooten00}
			E. L. Wooten, K. M. Kissa, A. Yi-Yan, E. J. Murphy, D. A. Lafaw, P. F. Hallemeier, D. Maack, D. V. Attanasio, D. J. Fritz, G. J. McBrien, and D. E. Bossi, ``A review of lithium niobate modulators for
			fiber-optic communications Systems," IEEE J. Sel. Top. Quant. Electron. {\bf 6}, 69 (2000).
			\bibitem{Capmany19}
			D. Marpaung, J. Yao, and J. Capmany, ``Integrated microwave photonics," Nature Photon. {\bf 13}, 80 (2019).
			\bibitem{Vladimir15}	
			C. Sun, M. T. Wade, Y. Lee, J. S. Orcutt, L. Alloatti, M. S. Georgas, A. S. Waterman, J. M. Shainline, R. R. Avizienis, S. Lin, B. R. Moss, R. Kumar, F. Pavanello, A. H. Atabaki, H. M. Cook, A. J. Ou, J. C. Leu, Y.-H. Chen, K. Asanovi\'{c}, R. J. Ram, M. A. Popovi\'{c}, and V. M. Stojanovi\'{c}, ``Single-chip microprocessor that communicates directly using light," Nature {\bf 528}, 534-538 (2015).
			\bibitem{Diddams18}
			D. R. Carlson, D. D. Hickstein, W. Zhang, A. J. Metcalf, F. Quinlan, S. A. Diddams, S. B. Papp, ``Ultrafast electro-optic light with subcycle control," Science {\bf 361}, 1358 (2018).
			\bibitem{Smith17}
			M. Karpi\'{n}ski, M. Jachura, L. J. Wright, and B. J. Smith, ``Bandwidth manipulation of quantum light by an electro-optic time lens," Nature Photon. {\bf 11}, 53 (2017).
			\bibitem{Reed10}
			G. T. Reed, G. Mashanovich, F. Y. Gardes, and D. J. Thomson, ``Silicon optical modulators," Nature Photon. {\bf 4}, 518 (2010).
			\bibitem{Keeler18}
			M. G. Wood, S. Campione, S. Parameswaran, T. S. LUK, J. R. WENDT, D. K. Serkland, and G. A. Keeler, ``Gigahertz speed operation of epsilon-near-zero silicon photonic modulators," Optica {\bf 2}, 233 (2018).
			\bibitem{Nakano18}
			J. Ozaki, Y. Ogiso, and S. Nakano, ``High-speed modulator for next-generation large-capacity coherent optical networks," NTT Technical Reviews {\bf 16}(4), 1 (2018).
			\bibitem{Michel08}
			J. Liu, M. Beals, A. Pomerene, S. Bernardis, R. Sun, J. Cheng, L. C. Kimerling, and J. Michel, ``Waveguide-integrated, ultralow-energy GeSi electro-absorption modulators," Nature Photon. {\bf 2}, 433 (2008).
			\bibitem{Koos15}
			S. Koeber, \emph{et al}, ``Femtojoule electro-optic modulation using a silicon–organic hybrid device," Light: Sci. Appl. {\bf 4}, e255 (2015).
			
			\bibitem{Gunter12}
			G. Poberaj, H. Hu, W. Sohler, and P. G\"{u}nter, ``Lithium niobate on insulator (LNOI) for micro-photonic devices," Laser Photon. Rev. {\bf 6}, 488 (2012).
			\bibitem{Bowers18}
			A. Boes, B. Corcoran, L. Chang, J. Bowers, and A. Mitchell, ``Status and Potential of Lithium Niobate on Insulator (LNOI) for Photonic Integrated Circuits," Laser Photon. Rev. {\bf 12}, 1700256 (2018).
			
			\bibitem{Gunter07}
			A. Guarino, G. Poberaj, D. Rezzonico, R. Gegl'Innocenti, and P. G\"{u}nter, ``Electro–optically tunable microring resonators in lithium niobate," Nature Photon. {\bf 1}, 407 (2007).
			\bibitem{Reano13}
			L. Chen, M. G. Wood, and R. M. Reano, ``12.5 pm/V hybrid silicon and lithium niobate optical microring resonator with integrated electrodes," Opt. Express {\bf 21}, 27003 (2013).
			\bibitem{Reano14}
			L. Chen, Q. Xu, M. G. Wood, and R. M. Reano, ``Hybrid silicon and lithium niobate electro-optical ring modulator," Optica {\bf 1}, 112 (2014).
			\bibitem{Fathpour15}
			A. Rao, A. Patil, J. Chiles, M. Malinowski, S. Novak, K. Richardson, P. Rabiei, and S. Fathpour, ``Heterogeneous microring and Mach-Zehnder modulators based on lithium niobate and chalcogenide glasses on silicon," Opt. Express {\bf 23}, 22746 (2015).
			\bibitem{Fathpour16}
			A. Rao, A. Patil, P. Rabiei, A. Honardoost, R. Desalvo, A. Paolella, and S. Fathpour, ``High-performance and linear thin-film lithium niobate Mach–Zehnder modulators on silicon up to 50 GHz," Opt. Lett. {\bf 41}, 5700 (2016).
			\bibitem{Amir17}
			J. D. Witmer, J. A. Valery, P. Arrangoiz-Arriola, C. J. Sarabalis, J. T. Hill, and A. H. Safavi-Naeini, ``High-Q photonic resonators and electro-optic coupling using silicon-on-lithium-niobate," Sci. Rep. {\bf 7}, 46313 (2017).
			\bibitem{Loncar18}		
			C. Wang, M. Zhang, B. Stern, M. Lipson, and M. Loncar, ``Nanophotonic lithium niobate electro-optic modulators," Opt. Express {\textbf 26}, 1547-1555 (2018).
			\bibitem{Prather18}
			A. J. Mercante, S. Shi, P. Yao, L. Xie, R. M. Weikle, and D. W. Prather, ``Thin film lithium niobate electro-optic modulator with terahertz operating bandwidth," Opt. Express {\bf 26}, 14810 (2018).
			\bibitem{Loncar18_2}
			C. Wang, M. Zhang, X. Chen, M. Bertrand, A. Shams-Ansari, S. Chandrasekhar, P. Winzer, and M. Loncar, ``Integrated lithium niobate electro-optic modulators operating at CMOS-compatible voltages," Nature {\bf 562}, 101 (2018).
			\bibitem{Shayan18}
			P. O. Weigel, \emph{et al}, ``Bonded thin film lithium niobate modulator on a silicon photonics platform exceeding 100 GHz 3-dB electrical modulation bandwidth," Opt. Express {\bf 26}, 23728 (2018).
			\bibitem{Fathpour18}
			A. Rao and S. Fathpour, ``Compact lithium niobate electrooptic modulators," IEEE J. Sel. Top. Quant. Electron. {\bf 24}, 3400114 (2018).
			\bibitem{Cai19}
			M. He, \emph{et al}, ``High-performance hybrid silicon and lithium niobate Mach Zehnder modulators for 100 Gbit s$^{-1}$ and beyond," Nature Photon. {\bf 13}, 359 (2019).
			\bibitem{Cai19_2}
			J. Jian, M. Xu, L. Liu, Y. Luo, J. Zhang, L. Liu, L. Zhou, H. Chen, S. Yu, and X. Cai, ``High modulation efficiency lithium niobate Michelson interferometer modulator," Opt. Express {\bf 27}, 18731 (2019).
			\bibitem{Loncar19}		
			M. Zhang, B. Buscaino, C. Wang, A. S. Ansari, C. Reimer, R. Zhu, J. Kahn, and M. Loncar, ``Broadband electro-optic frequency comb generation in a lithium niobate microring resonator," Nature {\bf 568} 373-377 (2019).
			
			\bibitem{Miller17}		
			D. A. B. Miller, ``Attojoule optoelectronics for low-energy information processing and communications," J. Lightwave Technol. {\bf 35}, 346-396 (2017).
			\bibitem{Sorger15}
			K. Liu, C. R. Ye, S. Khan, and V. J. Sorger, ``Review and perspective on ultrafast wavelength-size electro-optic modulators," Laser. Photon. Rev. {\bf 9}, 172 (2015).
			
			
			\bibitem{Notomi09}			
			Takasumi Tanabe, Katsuhiko Nishiguchi, Eiichi Kuramochi, and Masaya Notomi, "Low power and fast electro-optic silicon modulator with lateral p-i-n embedded photonic crystal nanocavity," Opt. Express 17, 22505-22513 (2009).
			\bibitem{Baba12}
			H. C. Nguyen, S. Hashimoto, M. Shinkawa, and T. Baba, ``Compact and fast photonic crystal silicon optical modulators," Opt. Express {\bf 20}, 22465 (2012).
			\bibitem{Notomi14}		
			Abdul Shakoor, Kengo Nozaki, Eiichi Kuramochi, Katsuhiko Nishiguchi, Akihiko Shinya, and Masaya Notomi, "Compact 1D-silicon photonic crystal electro-optic modulator operating with ultra-low switching voltage and energy," Opt. Express 22, 28623-28634 (2014)
			\bibitem{Jelena11}		
			Gary Shambat, Bryan Ellis, Marie A. Mayer, Arka Majumdar, Eugene E. Haller, and Jelena Vuckovic, "Ultra-low power fiber-coupled gallium arsenide photonic crystal cavity electro-optic modulator," Opt. Express 19, 7530-7536 (2011)
			\bibitem{Notomi19}
			Kengo Nozaki, Shinji Matsuo, Takuro Fujii, Koji Takeda, Akihiko Shinya, Eiichi Kuramochi, and Masaya Notomi, ``Femtofarad optoelectronic integration demonstrating energy-saving signal conversion and nonlinear functions",  Nat. Photonics {\bf 13}, 454â€“459 (2019)
			
			\bibitem{Krauss09}
			J. H. W\"{u}lbern, J. Hampe, A. Petrov, M. Eich, J. Luo, A. K.-Y. Jen, A. Di Falco, T. F. Krauss, and J. Bruns, ``Electro-optic modulation in slotted resonant photonic crystal heterostructures," Appl. Phys. Lett. {\bf 94}, 241107 (2009).
			\bibitem{Chen16}
			X. Zhang, C.-J. Chung, A. Hosseini, H. Subbaraman, J. Luo, A. K-Y. Jen, R. L. Nelson, C. Y-C. Lee, and R. T. Chen, ``High performance optical modulator based on electro-optic polymer filled silicon slot photonic crystal waveguide," J. Lightwave Technol. {\bf 34}, 2941 (2016).
			\bibitem{AlanX19}		
			E. Li, B. Zhou, Y. Bo, and A. X. Wang, "High-Speed Compact Silicon Nanocavity Modulator with Transparent Conductive Oxide Gate," in Frontiers in Optics + Laser Science APS/DLS, OSA Technical Digest (Optical Society of America, 2019), paper FW5C.2.
			
			\bibitem{Bernal12}
			H. Lu, F. I. Baida, G. Ulliac, N. Courjal, M. Collet, and M.-P. Bernal, ``Lithium niobate photonic crystal wire cavity: Realization of a compact electro-optically tunable filter,'' Appl. Phys. Lett. {\bf 101}, 151117 (2012).
			\bibitem{Bernal12_3}
			H. Lu, B. Sadani, N. Courjal, G. Ulliac, N. Smith, V. Stenger, M. Collet, F. I. Baida, and M.-P. Bernal, ``Enhanced electro-optic lithium niobate photonic crystal wire waveguide on a smart-cut thin film," Opt. Express {\bf 20}, 2974 (2012).
			\bibitem{Bernal14}
			H. Lu, W. Qiu, C. Guyot, G. Ulliac, J.-M. Merolla, F. Baida, and M.-P. Bernal, ``Optical and RF characterization of a lithium niobate photonic crystal modulator," IEEE Photon. Technol. Lett. {\bf 26}, 1332 (2014).
			
			\bibitem{Liang17}
			H. Liang, R. Luo, Y. He, H. Jiang, and Q. Lin, ``High-quality lithium niobate photonic crystal nanocavities," Optica {\bf 4}, 1251 (2017).
			\bibitem{Li19}
			M. Li, H. Liang, R. Luo, Y. He, and Q. Lin, ``High‐Q 2D Lithium Niobate Photonic Crystal Slab Nanoresonators," Laser Photon. Rev. {\bf 13}, 1800228 (2019).
			\bibitem{Amir19}			
			W. Jiang, R. N. Patel, F. M. Mayor, T. P. McKenna, P. Arrangoiz-Arriola, C. J. Sarabalis, J. D. Witmer, R. Van Laer, and A. H. Safavi-Naeini, ``Lithium niobate piezo-optomechanical crystals," Optica {\bf 6}, 845-853 (2019).
			\bibitem{Mingxiao192}		
			M. Li, H. Liang, R. Luo, Y. He, J. Ling, and Q. Lin, ``Photon-level tuning of photonic nanocavities,"Optica {\bf 6}, 860-863 (2019).
			
			
			\bibitem{Usman19}	
			U. A. Javid and Q. Lin, ``Quantum correlations from dynamically modulated optical nonlinear interactions", Phys. Rev. A, 100, 043811 (2019).
			
			\bibitem{Piazza19}
			L. Cai, A. Mahmoud, M. Khan, M. Mahmoud, T. Mukherjee, J. Bain, and G. Piazza, ``Acousto-optical modulation of thin film lithium niobate waveguide devices," Photon. Res. {\bf 7}, 1003 (2019).
			\bibitem{Loncar196}
			L. Shao, M. Yu, S. Maity, N. Sinclair, L. Zheng, C. Chia, A. Shams-Ansari, C. Wang, M. Zhang, K. Lai, and M. Loncar, ``Microwave-to-optical conversion using lithium niobate thin-film acoustic resonators," Optica {\bf 6}, 1498 (2019).
			
			
			
			
			
			
			
			
			
			
			
			
			
			
			
			
			
			
			
			
			
			
			
			
			
			
			
			
			
			
			
			
			
			
			
			
			
			
			
			
			
			
			
			
			
		\end{thebibliography}
\end{document}